\documentclass[conference]{IEEEtran}
\IEEEoverridecommandlockouts
\usepackage{amsmath}  
\usepackage{amssymb}                
\usepackage{graphicx}               
\usepackage{array}  
\usepackage{tabularx}
\usepackage{adjustbox}
\usepackage[utf8]{inputenc}
\usepackage[T1]{fontenc}
\usepackage{epsfig,endnotes}
\usepackage{multirow}
\usepackage{booktabs}
\usepackage{graphicx}
\usepackage{amsfonts}
\usepackage{cite}
\usepackage{textcomp}
\usepackage{xcolor}
\usepackage{algorithm}
\usepackage{algorithmic}

\usepackage{float}
\usepackage{makecell}

\def\BibTeX{{\rm B\kern-.05em{\sc i\kern-.025em b}\kern-.08em
    T\kern-.1667em\lower.7ex\hbox{E}\kern-.125emX}}

\makeatletter
\newcommand{\linebreakand}{%
  \end{@IEEEauthorhalign}
  \hfill\mbox{}\par
  \mbox{}\hfill\begin{@IEEEauthorhalign}
}
\makeatother

\usepackage{etoolbox}
\makeatletter
\patchcmd{\@makecaption}
  {\scshape}{}{}{}
\patchcmd{\@makecaption}
  {\MakeUppercase}{}{}{}
\makeatother

\begin{document}

\title{WebGuard++:Interpretable Malicious URL Detection via Bidirectional Fusion of HTML Subgraphs and Multi-Scale Convolutional BERT}

\author{\IEEEauthorblockN{1\textsuperscript{st} Ye Tian}
\IEEEauthorblockA{\textit{Hangzhou Research Institute} \\
\textit{Xidian University}\\
Hangzhou, China\\
tianye@xidian.edu.cn}
\and
\IEEEauthorblockN{2\textsuperscript{nd} ZhangYumin}
\IEEEauthorblockA{\textit{Hangzhou Research Institute} \\
\textit{Xidian University}\\
Hangzhou, China\\
24241214904@stu.xidian.edu.cn}
\and
\IEEEauthorblockN{3\textsuperscript{rd} Yifan Jia}
\IEEEauthorblockA{\textit{Yantai Research Institute} \\
\textit{Harbin Engineering University}\\
Yantai, China \\
jiayf@hrbeu.edu.cn}
\linebreakand 
\IEEEauthorblockN{4\textsuperscript{th} Jianguo Sun\textsuperscript{*}}
\IEEEauthorblockA{\textit{Hangzhou Research Institute} \\
\textit{Xidian University}\\
Hangzhou, China\\
jgsun@xidian.edu.cn}
\and
\IEEEauthorblockN{5\textsuperscript{th} Yanbin Wang\textsuperscript{*}}
\IEEEauthorblockA{\textit{Hangzhou Research Institute} \\
\textit{Xidian University}\\
Hangzhou, China\\
wangyanbin15@mails.ucas.ac.cn}

\thanks{\textsuperscript{*}Yanbin Wang and Jianguo Sun are co-corresponding authors.}
}


\maketitle

\begin{abstract}
URL+HTML feature fusion shows promise for robust malicious URL detection, since attacker artifacts persist in DOM structures. However, prior work suffers from four critical shortcomings: (1) incomplete URL modeling, failing to jointly capture lexical patterns and semantic context; (2) HTML graph sparsity, where threat-indicative nodes (e.g., obfuscated scripts) are isolated amid benign content, causing signal dilution during graph aggregation; (3) unidirectional analysis, ignoring URL-HTML feature bidirectional interaction; and (4) opaque decisions, lacking attribution to malicious DOM components.

To address these challenges, we present WebGuard++, a detection framework with 4 novel components: 1) Cross-scale URL Encoder: Hierarchically learns local-to-global and coarse to fine URL features based on Transformer network with dynamic convolution. 2) Subgraph-aware HTML Encoder: Decomposes DOM graphs into interpretable substructures, amplifying sparse threat signals via Hierarchical feature fusion. 3) Bidirectional Coupling Module: Aligns URL and HTML embeddings through cross-modal contrastive learning, optimizing inter-modal consistency and intra-modal specificity. 4) Voting Module: Localizes malicious regions through consensus voting on malicious subgraph predictions. Experiments show WebGuard++ achieves significant improvements over state-of-the-art baselines,  achieving 1.1×–7.9× higher TPR at fixed FPR of 0.001 and 0.0001 across both datasets.

\end{abstract}


\IEEEpeerreviewmaketitle

\section{Introduction}
Phishing attacks have become one of the most pervasive and damaging cyber threats in recent years, with the Anti-Phishing Working Group (APWG) reporting record-breaking volumes of 1,624,144 attacks in Q1 2023~\cite{ALJOFEY2025104170} and approximately one million in Q1 2024~\cite{ZHANG2025111303}. Modern phishing campaigns employ increasingly sophisticated evasion techniques, including homoglyphic domain spoofing (e.g., "facbook.com"), URL shortening services, and malicious content embedded within legitimate platforms. These deceptive practices enable attackers to bypass traditional URL detection methods, such as blacklists~\cite{10.1145/1456424.1456434, AZEEZ2021102328}, rules~\cite{MOGHIMI2016231} and manual feature engineering~\cite{WAZIRALI2021108591}, while effectively mimicking trusted websites to steal sensitive information. The growing sophistication and scale of these threats underscore the critical need for robust, accurate, and efficient phishing detection systems capable of operating at web scale with near-zero false positive rates to adequately protect users and infrastructure.

Most approaches to malicious URL detection have predominantly relied on URL-based features~\cite{sahoo2019maliciousurldetectionusing}. However, URLs present several critical limitations: 1) Limited Information Dimensions – URLs encode minimal contextual data, restricting detection models to surface-level patterns. 2) Vulnerability to Evasion – Attackers easily manipulate URLs through obfuscation, mimicry, or rapid replacement to bypass detection. 3) Lack of Structural Insight – URLs alone cannot reveal the underlying page behaviors that indicate malicious intent.

Integrating HTML structural analysis has the potential to address these shortcomings. Unlike URLs, HTMLs contain: 1) Rich Hierarchical Structure – DOM trees, nested iframes, and script dependencies expose hidden attack vectors (e.g., phishing content loaded via iframes). 2) Interaction Logic – Form actions, redirects, and domain mismatches reveal data exfiltration attempts (e.g., spoofed submission endpoints). By training models to recognize these structural anomalies, HTML+URL detection complements and reinforces URL-based features and mitigates evasion tactics, as structural manipulations are harder to disguise than URL alterations. 

Existing approaches like Web2Vec and PhishDet demonstrate progressive advancements. Web2Vec proposes a deep hybrid network architecture that jointly processes URL strings, HTML content, and DOM structures. PhishDet combines Long Short-Term Memory (LSTM) networks with Graph Convolutional Networks (GCNs) to model URL patterns and structural HTML features. However, these methods exhibit 4 critical limitations: 
\begin{itemize}
    \item Incomplete URL Modeling: Existing methods process URLs as either lexical sequences or static tokens, failing to capture (i) local character-level manipulations (e.g., "facwook.com") and (ii) global semantic context (e.g. deceptive subdomains).
    \item HTML Graph Sparsity: Current GNN-based DOM analyzers suffer from signal dilution during neighborhood aggregation, as threat-indicative nodes (e.g., obfuscated <script> tags) account for <5\% of typical DOM graphs, while benign content dominates attention weights.
    \item Deficient Cross-Modal Interaction: Existing methods have not modeled dynamic, bidirectional semantic relationships between modalities, missing mutually reinforcing signals. For instance, a suspicious URL (e.g., "login.paypa1.com") may align with HTML structures like payment forms, while anomalous DOM elements (e.g., fake brand-related notices) can clarify URL intent.
    \item Opaque Decisions: Black-box architectures lack attribution mechanisms to pinpoint malicious DOM components, hindering forensic analysis.
\end{itemize}

We propose WebGuard++ to overcome cross-modal phishing detection challenges with four interlocking technical: (1) Cross-scale URL Encoder: Combines ConvBERT's hierarchical representations with spatial pyramid fusion to jointly capture character-level obfuscational patterns (e.g., "paypa1") and semantic inconsistencies (e.g., deceptive subdomains). (2) Subgraph-aware HTML Encoder: Performs DOM subgraph partitioning with stabilized node grouping and iterative batch sampling, enabling localized malicious signal aggregation (e.g., form clusters) while mitigating benign node interference in full-graph processing. (3) Bidirectional Coupling Module: Employs stacked feature layers with both self and cross-attention to capture distinct semantic subspaces, enabling bidirectional URL-HTML feature interaction. (4) Voting Module: Adopts a minimum compromise voting policy - any malicious subgraph ( $\geq 1$ in sampled rounds) triggers global malicious classification, while providing actionable forensic evidence through malicious subgraphs. 

This work makes the following key contributions:

\begin{itemize}
    \item We propose a cross-modal malicious URL detection that achieves 1.1-7.9× higher TPR at <0.01 FPR by jointly modeling URL and HTML features.
    \item We design a URL encoding method that captures both lexical obfuscation patterns and semantic inconsistencies by multi-layer ConvBERT with pyramidal feature fusion.
    \item We propose a subgraph learning method for HTML that employs effective subgraph partitioning to aggregate local malicious signals while preventing feature dilution.
    \item We use a hybrid attention network with both self/cross-attention to learn bidirectional, multi-view relationships between URLs and HTML content.
    \item Our method is the first subgraph-based maliciousness prediction that provides both fine-grained classification and component-level traceability.
\end{itemize}

\begin{figure*}[htbp!]  
    \centering
    \includegraphics[width=0.98\textwidth]{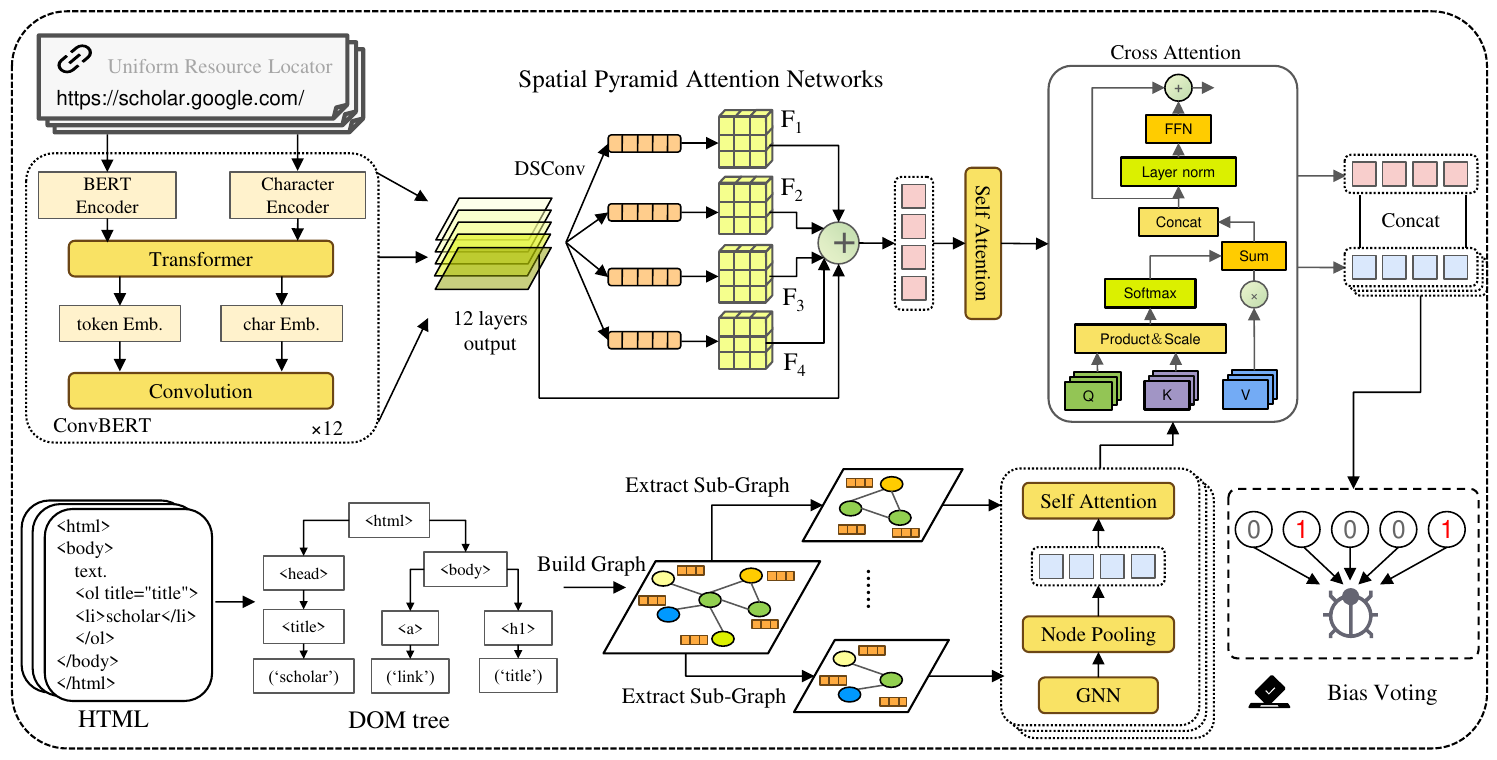}  
    \caption{Framework diagram of the model structure of WebGuard++.}  
    \label{fig:WebGuard++}  
\end{figure*}

\section{Related Work}
Early phishing detection methods mainly rely on extracting discriminative features from raw URLs. PhishDef~\cite{le2011phishdef} demonstrated that phishing links could be effectively identified using only static URL features, such as domain length and character composition, thereby reducing reliance on external resources. Nonetheless, it showed limited effectiveness against semantically natural and structurally sophisticated malicious URLs. PhishZoo~\cite{afroz2011phishzoo} enhanced the ability to detect spoofed pages by analyzing visual similarities between webpages, but its stability degraded under dynamic content or slight layout changes. Whereupon, Sahingoz et al~\cite{sahingoz2019machine} modeled URL character features using various classifiers, significantly improving detection accuracy and generalization capabilities.
Shraddha Parekh et al.~\cite{8473085} put forth a model by using the URL detection method using Random Forest algorithm. However, there are still some defects, such as the lack of fine-grained feature acquisition of URL text.

Despite these advancements, single-modality detection methods~\cite{tsai2024generalizedmaliciousurldetection} remain vulnerable to adaptive attacks that exploit their limited perceptual scope. With the advent of deep learning, more expressive models emerged. URLNet~\cite{le2018urlnet} combined character-level and word-level embeddings via convolutional neural networks (CNNs) to capture morphological patterns in URLs, offering significant improvements over handcrafted features. This line of research was extended through CNN-based architectures~\cite{alsadig2024CNN-based} and attention-based transformers such as TransURL~\cite{liu2024transurl}, which demonstrated stronger robustness against adversarial obfuscation. Concurrently, models such as PhishGuard~\cite{islam2024phishguard} and Fed-urlBERT~\cite{li2023fedurlbertclientsidelightweightfederated} integrated federated training and transformer encoders to support privacy-preserving and scalable learning. PhishBERT~\cite{10095719} further explored pre-trained language models for URL representation learning, yielding enhanced generalization.

In response to the continuously evolving phishing techniques, researchers have gradually shifted towards multimodal fusion and structure-aware modeling~\cite{299740,9415016,10976643,299838,5975221,10.5555/2887007.2887140,9582835,10179461,Guo_Cho_Chen_Sengupta_Hong_Mitra_2022}. Yoon et al.~\cite{yoon2024phishing} proposed a detection framework that integrates HTML DOM graphs and URL features based on graph convolutional and transformer networks. Lee et al.~\cite{lee2024multimodal} introduced a brand consistency verification mechanism, effectively enhancing performance under adversarial attacks, although challenges remain in identifying emerging niche brands. PhishAgent~\cite{cao2025phishagent} further improved detection robustness by leveraging a multimodal large language model to integrate webpage text, visual, and structural information. Lihui Meng et al.~\cite{10928669} proposes DPMLF (Deep Learning Phishing Detection Model with Multi-Level Features), which integrates URL character-level and HTML word-level semantic features. However, feature fusion using fully connected layers cannot closely match the modality, thus affecting the model performance.

In parallel, PhishIntention~\cite{10976643} and Phishpedia~\cite{9415016} emphasized visual understanding. A broader evaluation by Bushra Sabir et al.\cite{9937071} revealed that many state-of-the-art models exhibit drastic performance degradation when facing adversarial URL samples, highlighting the fragility of current systems.

Our approach fundamentally advances malicious URL detection by differentiating itself from prior work through four key dimensions: (1) fine-grained URL feature extraction, (2) subgraph-level HTML structure learning, (3) bidirectional modal coupling, and (4) malicious segment localization. 

\section{Methodology}

We organize this section as follows: First introducing data preprocessing, then detailing three core components (URL encoder, HTML encoder, and their bidirectional coupling), and finally presenting the phishing detection mechanism. See Figure~\ref{fig:WebGuard++} for overview.

\subsection{Cross-scale URL Encoder}
URLs serve as a fundamental indicator for malicious webpage detection, as attackers frequently manipulate URL structures to mimic benign pages while evading traditional pattern-matching techniques and retaining malicious intent. However, extracting discriminative URL features requires careful consideration of two key challenges: (1) structural ambiguity; and (2) adversarial noise—embedded homoglyphs or Base64-encoded payloads. This necessitates fine-grained URL feature extraction capable of capturing multiscale information at both character-level and semantic levels.

To address this, we integrate CharBERT and ConvBERT models, where the former extracts character-level URL features, while the latter, equipped with global and local context learning, refines local and global semantic representations. Furthermore, to enable coarse-to-fine representation learning, we extract embeddings from all 12 hidden layers of the ConvBERT model to construct a feature matrix, which is then processed via a spatial pyramid multiscale feature fusion module for hierarchical feature learning.

The spatial pyramid module first applies DSConv3×3 (depthwise separable convolution) to preliminarily transform input features, as follows:
\begin{equation}
X = DSConv3\times3(x)
\end{equation}

where the input is x and the output is X. Next, multiple DSConv3x3 convolutional kernel branches with different expansion rates are defined, where different expansion rates can capture contextual information at different scales. Smaller expansion rates can focus on local features, while larger expansion rates can capture more global features.
Each branch uses the depth-separable convolution DSConv3x3 with expansions $d_1$, $d_2$, $d_3$, $d_4$.
\begin{equation}
D_1 = DSConv3\times3(X,d_i),i=1,2,3,4
\end{equation}

where $D_i$ denotes the output of the $i^{th}$ branch and $d_i$ is the expansion rate of the branch.

Sum the outputs of all branches with the output of the initial convolution to achieve feature fusion.
\begin{equation}
DX = X + \sum_{i=1}^{4} D_i
\end{equation}

Next, along the lines of Spatial Pyramid Attention Networks~\cite{10.1007/978-3-030-58589-1_19}, pooling operations are performed on the feature maps using different sizes of Adaptive Average Pooling Layers (AdaptiveAvgPool2d) to extract features at different spatial scales.
\begin{equation}
y_1 = AdaptiveAvgPool2d(1)(DX)
\end{equation}
\begin{equation}
y_2 = AdaptiveAvgPool2d(2)(DX)
\end{equation}
\begin{equation}
y_3 = AdaptiveAvgPool2d(4)(DX)
\end{equation}

where AdaptiveAvgPool2d(k) denotes the adaptive pooling of the feature map to a size of $k\times k$.

Finally, these feature vectors are spliced and input to the fully connected layer for feature fusion, and the attention weights are obtained using the sigmoid function.
\begin{equation}
y = concat(y_1,y_2,y_3)
\end{equation}
\begin{equation}
y = FC_2(ReLU(FC_1(y)))
\end{equation}
\begin{equation}
y = sigmoid(y)
\end{equation}

where $FC_1$ and $FC_2$ denote the fully connected layers, respectively.

The attention weights are multiplied with the fused feature map to obtain the attention weighted feature map. The attention weighted feature map is summed with the original input feature map to realize the residual join to preserve the information of the original input.
\begin{equation}
O_{url} = DX \times y + x
\end{equation}

\begin{algorithm}[htbp]
\caption{Biased voting mechanism process.}
\label{alg:subgraph_voting}
\begin{algorithmic}[1]
\STATE $subgraphs \leftarrow sum(division\_func(graph))$
\STATE 0count $\leftarrow 0$
\STATE 1count $\leftarrow 0$
\STATE 0scores $\leftarrow 1$
\STATE 1scores $\leftarrow 0$
\FOR{$0$ to $iters\_per - 1$}
    \STATE selected\_subgraphs $\leftarrow random.sample(subgraphs, 4)$
    \STATE outputs $\leftarrow model(selected\_subgraphs, *url)$
    \STATE scores $\leftarrow \text{softmax}(outputs)$
    \STATE \_, predicted\_classes $\leftarrow \text{max}(outputs, 1)$
    \IF{$predicted == 0$}
        \STATE $0count \leftarrow 0count + 1$
        \STATE $0scores \leftarrow \min(0scores, scores)$
    \ELSE
        \STATE $1count \leftarrow 1count + 1$
        \STATE $1scores \leftarrow \max(1scores, scores)$
    \ENDIF
\ENDFOR
\STATE $y\_pred \leftarrow []$
\STATE $y\_scores \leftarrow []$
\IF{$1count \ge 2$}
    \STATE $1$ add to $y\_pred$
    \STATE $1scores$ add to $y\_scores$
\ELSE
    \STATE $0$ add to $y\_pred$
    \STATE $0scores$ add to $y\_scores$
\ENDIF

\textbf{return:} $y\_pred, y\_scores$
\end{algorithmic}
\end{algorithm}

\subsection{Subgraph-aware HTML Encoder}
The input HTML document is parsed using the Beautiful Soup library (v4.12.3) to construct a Document Object Model (DOM) tree representation. This tree preserves the hierarchical structure of HTML elements, including all tags, attributes, and text content. The DOM tree is traversed using a depth-first search (DFS) algorithm to generate: A diagram object for the NetworkX package (v3.1) where: Nodes represent HTML elements (tags) with unique identifiers. Edges encode parent-child relationships between elements.
A node attribute list containing structured metadata for each node: Tag type (e.g., <div>, <a>), Key-value pairs of HTML attributes (e.g., \{"class": "container"\}), Raw text content. Node features consist of node "tag", "attributes", "text" attributes text content. For each node, we use a pretrained Word2Vec model (minimum vocabulary size) to generate a 100-dimensional embedding.

Then, the corresponding edge matrices, node neighbor lists, and maximum number of neighbors are computed based on the edge and node data of the networkx.DiGraph() graph object, and all the data are integrated into the model-acceptable S2VGraph object structure. 

For HTML content, our objective is to learn effective representations of malicious signals embedded within its structure. However, since such signals are often concealed within normal HTML elements, they tend to be obscured during graph-based learning. To address this challenge, we propose subgraph-aware HTML graph learning, which transforms the HTML DOM into a graph structure and employs node-level subgraph partitioning to extract localized subgraphs. This approach enables subgraph-level feature learning, ensuring that malicious signals remain distinguishable and are not diluted by benign structural patterns.

Specifically, we partition the graph into N subgraphs using a hash function $H$. The function takes a node ID string ( $S_v$ ) as input and assigns nodes to distinct subsets, where each subgraph retains only the nodes and edges belonging to a specific group.

\begin{equation}
X^t = \{ X_v|H(S_v)\% T_f + 1 = t \} , t=1,2,...,T_f
\end{equation}

Where G=(V,E,X), G denotes the input graph, V denotes the node union, E denotes the edge union, X denotes the node feature matrix and $X_v$ denotes the feature vector of node v.

Next, we will select batch subgraphs to input into the model many times for neighbor aggregation, MLP nonlinear transformation and BatchNorm operations.
\begin{equation}
X_{concat}=Concat(G_1.node_features,...,G_B.node_features)
\end{equation}
\begin{equation}
H_0 = X_{concat}
\end{equation}
\begin{equation}
H^{(l)} = A_{block} \cdot H^{(l-1)}
\end{equation}
\begin{equation}
H^{(l)} = ReLU(BatchNorm(MLP_l(H^{(l)})))
\end{equation}

Where B denotes the batch size, $G_1$ represents the $1^{st}$ subgraph, l denotes the $l^{th}$ propagation, $H^{(l)}$ denotes the hidden representation of each layer, $A^{block}$denotes the block diagonal sparse matrix.

Finally, all sub-graphs of each layer are pooled at the sub-graph level, and the features of each layer are concatenated to obtain the final features of these sub-graphs.

\begin{equation}
{H_{pooled}}^{(i)} = P\cdot H^{(i)},i=0,1,...l
\end{equation}

Where P represents the pooling operation.

\begin{figure}[t]  
    \centering
    \includegraphics[width=0.5\textwidth]{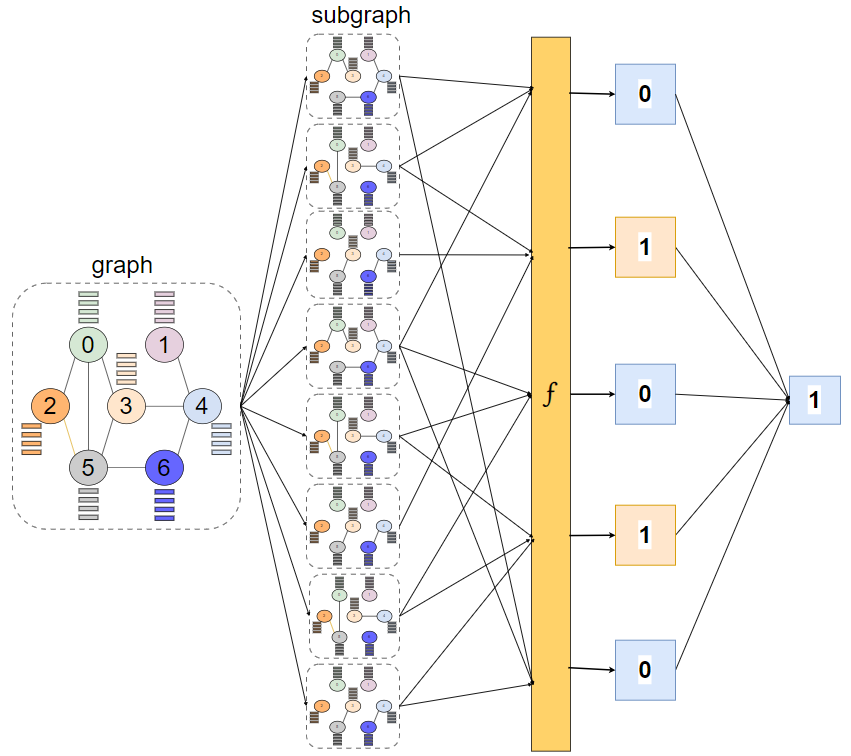}  
    \caption{Biased voting mechanism process.}  
    \label{fig:VotingMechanism}  
\end{figure}

\subsection{Bidirectional Coupling Module}
To learn joint URL-HTML representations, we propose bidirectional multi-view coupling (~\cite{10.1007/978-3-031-72970-6_3}) for fusing multimodal URL and HTML features. Unlike simple concatenation or unidirectional cross-attention layers, our bidirectional coupling module stacks multiple hybrid attention layers. Each layer combines self-attention and cross-attention mechanisms: self-attention independently models intra-modal dependencies within each modality, while cross-attention captures inter-modal interactions between URL and HTML features.

The HTML modality features are processed through self-attention to capture intra-modal relationships:
\begin{equation}
SelfAttention(Q_h,K_h,V_h) = softmax(\left( \frac{Q_hK^T_h}{\sqrt{d_k}} \right))V_h
\end{equation}

Similarly, URL features undergo self-attention:
\begin{equation}
SelfAttention(Q_u,K_u,V_u) = softmax(\left( \frac{Q_uK^T_u}{\sqrt{d_k}} \right))V_u
\end{equation}

We establish bidirectional cross-modal attention for inter-modal information exchange: 

HTML-to-URL attention extracts domain semantics from URL features:
\begin{equation}
CrossAttention(Q_u,K_h,V_h) = softmax(\left( \frac{Q_uK^T_h}{\sqrt{d_k}} \right))V_h
\end{equation}

URL-to-HTML attention captures structural patterns from HTML features:

\begin{equation}
CrossAttention(Q_h,K_u,V_u) = softmax(\left( \frac{Q_hK^T_u}{\sqrt{d_k}} \right))V_u
\end{equation}

The fused features are subsequently output.

\begin{table*}[t]
    \centering
    \caption{MTLP Dataset 10000 Evaluation metrics at scale: WebGuard++ vs other models.}
    \label{tab:10000_1}
    \resizebox{0.98\textwidth}{!} { 
    \fontsize{12pt}{16pt}\selectfont 
    \begin{tabular}{@{}lcccccccccccc@{}}
        \toprule
        \textbf{Method}
        & \textbf{TN} & \textbf{FP} & \textbf{FN} & \textbf{TP} & \textbf{ACC} & \textbf{Precision} & \textbf{Recall} & \textbf{F1} & \textbf{PR-AUC}
        & \textbf{ROC-AUC} & \textbf{MCC} & \textbf{Weighted F1}
         \\
        \midrule
        {\scshape BiLSTM}  & 940 & 54 & 922 & 84 & 0.5120 & 0.6086 
        & 0.0834 & 0.1468 & 0.5118
        & 0.5145 & 0.0575 & / \\
        {\scshape TextCNN}   & 814 & 180 & 104 & 902 & 0.8580 & 0.8336 
        & 0.8966 & 0.8639 & 0.9472
        & 0.9434 & 0.7179 & 0.8577
        \\
        {\scshape URLNet}   & 829 & 165 & 148 & 857 & 0.9269 & 0.8385 
        & 0.8527 & 0.8455 & 0.7891
        & 0.8433 & 0.6869 & 0.8434
        \\
        {\scshape TransURL}   & 953 & 41 & 20 & 986 & 0.9695 & 0.9600
        & 0.9801 & 0.9699 & 0.9934
        & 0.9937 & 0.9391 & 0.9694
        \\
        {\scshape PMANet} & 991 & 26 & 22 & 961 & 0.9760 & 0.9736 & 0.9776 & 0.9756 & 0.9933 & 0.9952 & 0.9519 & 0.9760 \\
        {\scshape URLBERT} & 910 & 42 & 11 & 1037 & 0.9735 & 0.9610 & \textbf{0.9895} & 0.9750 & \textbf{0.9965} & \textbf{0.9967} & 0.9472 & 0.9734 \\
        {\scshape dephides} & 966 & 36 & 76 & 922 & 0.9440 & 0.9624 & 0.9238 & 0.9427 & 0.9824 & 0.9827 & 0.8887 & 0.9439 \\
        {\scshape Semi-GAN} & 790 & 151 & 17 & 948 & 0.9118 & 0.8626 & 0.9823 & 09186 & / & 0.9826 & 0.8316 & 0.9113 \\

        \midrule
        \textbf{{\scshape WebGuard++}}      & 989 & 21 & 24 & 966 & \textbf{0.9775} & \textbf{0.9787} 
        & 0.9757 & \textbf{0.9772} & 0.9959
        & 0.9949 & \textbf{0.9549} & \textbf{0.9772}
        \\
        \bottomrule
    \end{tabular}
    }
\end{table*}
\begin{table*}[t]
    \centering
    \caption{The Abdelhakim Dataset phishing dataset Evaluation metrics : WebGuard++ vs other models.}
    \label{tab:A_1}
    \resizebox{0.98\textwidth}{!} { 
    \fontsize{12pt}{16pt}\selectfont 
    \begin{tabular}{@{}lcccccccccccc@{}}
        \toprule
        \textbf{Method}
        & \textbf{TN} & \textbf{FP} & \textbf{FN} & \textbf{TP} & \textbf{ACC} & \textbf{Precision} & \textbf{Recall} & \textbf{F1} & \textbf{PR-AUC}
        & \textbf{ROC-AUC} & \textbf{MCC} & \textbf{Weighted F1}
        \\
        \midrule
        {\scshape BiLSTM}  & 115 & 2 & 58 & 13 & 0.6808 & 0.8666 
        & 0.1830 & 0.3023 & 0.4671
        & 0.5830 & 0.2970 & / \\
        {\scshape TextCNN}   & 103 & 14 & 42 & 29 & 0.7021 & 0.6744 
        & 0.4084 & 0.5087 & 0.6462
        & 0.7647 & 0.3333 & 0.6814
        \\
        {\scshape URLNet}   & 90 & 26 & 40 & 31 & 0.8128 & 0.5438 
        & 0.4366 & 0.4843 & 0.4513
        & 0.6062 & 0.2240 & 0.6378
        \\
        {\scshape TransURL}   & 101 & 16 & 20 & 51 & 0.8085 & 0.7611 
        & 0.7183 & 0.7391 & 0.8210
        & 0.8737 & 0.5886 & 0.8073
        \\
        {\scshape PMANet} & 98 & 14 & 14 & 62 & 0.8510 & 0.8157 & 0.8157 & 0.8157 & 0.8827 & 0.8956 & 0.6907 & 0.8510 \\
        {\scshape URLBERT} & 88 & 18 & 16 & 66 & 0.8191 & 0.7857 & 0.8048 & 0.7951 & 0.8772 & 0.9004 & 0.6334 & 0.8193 \\
        {\scshape dephides} & 104 & 1 & 67 & 16 & 0.6382 & \textbf{0.9411} & 0.1927 & 0.3200 & 0.7898 & 0.7840 & 0.3172 & 0.5621 \\
        {\scshape Semi-GAN} & 88 & 24 & 34 & 35 & 0.6795 & 0.5932 & 0.5072 & 0.5468 & 0.4759 & 0.7017 & 0.3035 & 0.6738 \\
        \midrule
        \textbf{{\scshape WebGuard++}}      & 109 & 8 & 12 & 59 & \textbf{0.8936} & 0.8805 
        & \textbf{0.8309} & \textbf{0.8550} & \textbf{0.9072}
        & \textbf{0.9256} & \textbf{0.7719} & \textbf{0.8550}\\
        \bottomrule
    \end{tabular}
    }
\end{table*}

\subsection{phishing bias voting mechanism}
we propose a biased voting mechanism for phishing website detection. If the number of malicious predictions across multiple rounds of batch subgraph extraction exceeds one, we directly classify the corresponding HTML-URL pair as a phishing website. Unlike conventional approaches, our mechanism operates at the subgraph level, enabling not only final detection but also localization of malicious regions within the HTML structure. The voting process is illustrated in Figure~\ref{fig:VotingMechanism}, and the algorithmic workflow is detailed in Algorithm~\ref{alg:subgraph_voting}.

Specifically, for each HTML graph divided into num\_groups of subgraphs, we will perform iter\_per rounds to randomly extract iter\_num subgraphs and URL data corresponding to the current HTML graph from the subgraphs and input them into the WebGuard++ model for feature extraction. WebGuard++'s Subgraph-aware HTML Encoder will batch process the subgraph features without affecting each other. Subsequently, WebGuard++ fuses the output subgraph features from the Subgraph-aware HTML Encoder to represent the total features of the subgraph collection in this round of extraction.

Suppose the current is a phishing site, num\_group=5, iter\_per=5, iter\_num=4, a subgraph contains malicious. Then, 4 images are randomly selected inside 5 images, and the probability that 4 images contain malicious subgraphs is 80\%; the probability that none of the set of subgraphs of the extraction site contains malicious subgraphs in each of the 5 rounds of random selection is 0.032\%, which is close to zero.

\begin{equation}
\text{Contains malicious} = \frac{C(4, 3)}{C(5, 4)} =\frac{4}{5} = 80\% 
\end{equation}
\begin{equation}
\text{Without malicious} = 20\% 
\end{equation}
\begin{equation}
\text{All without malice} = (20\%)^5 = 0.032\% 
\end{equation}

Therefore, our method can basically extract subgraphs with full coverage and input them to the model for prediction.

To ensure the accuracy of the predictions, we set the condition for determining URL-HTML data pairs as malicious to be when the number of predictions as malicious in all extraction rounds is greater than 1. When a batch of subgraph features is predicted to be malicious, the current URL-HTML data pair is tagged as malicious. If a subsequent batch of subgraph features is also predicted to be malicious, the previous prediction is confirmed to be true, indicating that the HTML graph does indeed contain a subgraph with malicious content, and the current URL-HTML data pair is malicious.

\begin{table}[t]
    \centering
    \caption{MTLP Dataset 10000 in terms of TPR@FPR metrics: WebGuard++ vs other models.}
    \label{tab:10000_2}
    \fontsize{28pt}{32pt}\selectfont 
    \begin{adjustbox}{max width=\columnwidth} 
    \begin{tabular}{@{}lcccc@{}}
        \toprule
        \textbf{Method} 
        & \textbf{\makecell{TPR@FPR\\(0.0001)}} 
        & \textbf{\makecell{TPR@FPR\\(0.001)}} 
        & \textbf{\makecell{TPR@FPR\\(0.01)}} 
        & \textbf{\makecell{TPR@FPR\\(0.1)}} \\
        \midrule
        {\scshape BiLSTM} & 0& 0 & 0& 0.0834  \\
        {\scshape TextCNN} & 0 & 0.2345 & 0.4771
        & 0.8429 \\
        {\scshape URLNet} & 0 & 0 & 0& 0 \\
        {\scshape TransURL} & 0.3797 & 0.3797 & 0.8876
        & 0.9960  \\
        {\scshape PMANet} & 0.2177 & 0.2553 & 0.9369 & 0.9979 \\
        {\scshape URLBERT} & 0.4122 & 0.4122 & 0.9408 & \textbf{0.9990} \\
        {\scshape dephides} & 0.1002 & 0.1002 & 0.8086 & 0.9829 \\
        {\scshape Semi-GAN} & 0 & 0 & 0.7709 & 0.9544 \\
        \midrule
        \textbf{{\scshape WebGuard++}} & \textbf{0.6848} & \textbf{0.7939} & \textbf{0.9515}
        & 0.9929 \\
        \bottomrule
    \end{tabular}
    \end{adjustbox}
\end{table}

\begin{table}[t]
    \centering
    \caption{The Abdelhakim Dataset phishing dataset in terms of TPR@FPR metrics: WebGuard++ vs other models.}
    \label{tab:A_2}
    \fontsize{28pt}{32pt}\selectfont 
    \begin{adjustbox}{max width=\columnwidth} 
    \begin{tabular}{@{}lcccc@{}}
        \toprule
        \textbf{Method} 
        & \textbf{\makecell{TPR@FPR\\(0.0001)}} 
        & \textbf{\makecell{TPR@FPR\\(0.001)}} 
        & \textbf{\makecell{TPR@FPR\\(0.01)}} 
        & \textbf{\makecell{TPR@FPR\\(0.1)}} \\
        \midrule
        {\scshape BiLSTM} & 0& 0 & 0& 0.1830  \\
        {\scshape TextCNN} & 0 & 0 & 0.0704
        & 0.3521 \\
        {\scshape URLNet} & 0 & 0 & 0& 0\\
        {\scshape TransURL} & 0.1126 & 0.1126 & 0.3521
        & 0.6338 \\
        {\scshape PMANet} & 0.2105 & 0.2105 & 0.4605 & 0.7368\\
        {\scshape URLBERT}& 0.3414 & 0.3414 & 0.3536 & 0.6707\\
        {\scshape dephides}& 0.1927 & 0.1927 & 0.2771 & 0.5421\\
        {\scshape Semi-GAN} & 0 & 0 & 0 & 0.3623\\
        \midrule
        \textbf{{\scshape WebGuard++}} & \textbf{0.3802} & \textbf{0.3802} & \textbf{0.5492}
        & \textbf{0.8591}\\
        \bottomrule
    \end{tabular}
    \end{adjustbox}
\end{table}

\section{Experiments}
In this section, we conduct extensive experiments to evaluate the proposed method. First, we describe the experimental setup, including the dataset, evaluation metrics, environment and equipment. Then, we compare the phishing site detection performance of our proposed method with other state-of-the-art methods. Finally, we perform some ablation experiments, cross-dataset tests and model robustness tests.

\subsection{EXPERIMENTAL SETUP}

\begin{figure*}[htbp!]  
    \centering
    \includegraphics[width=0.95\linewidth]{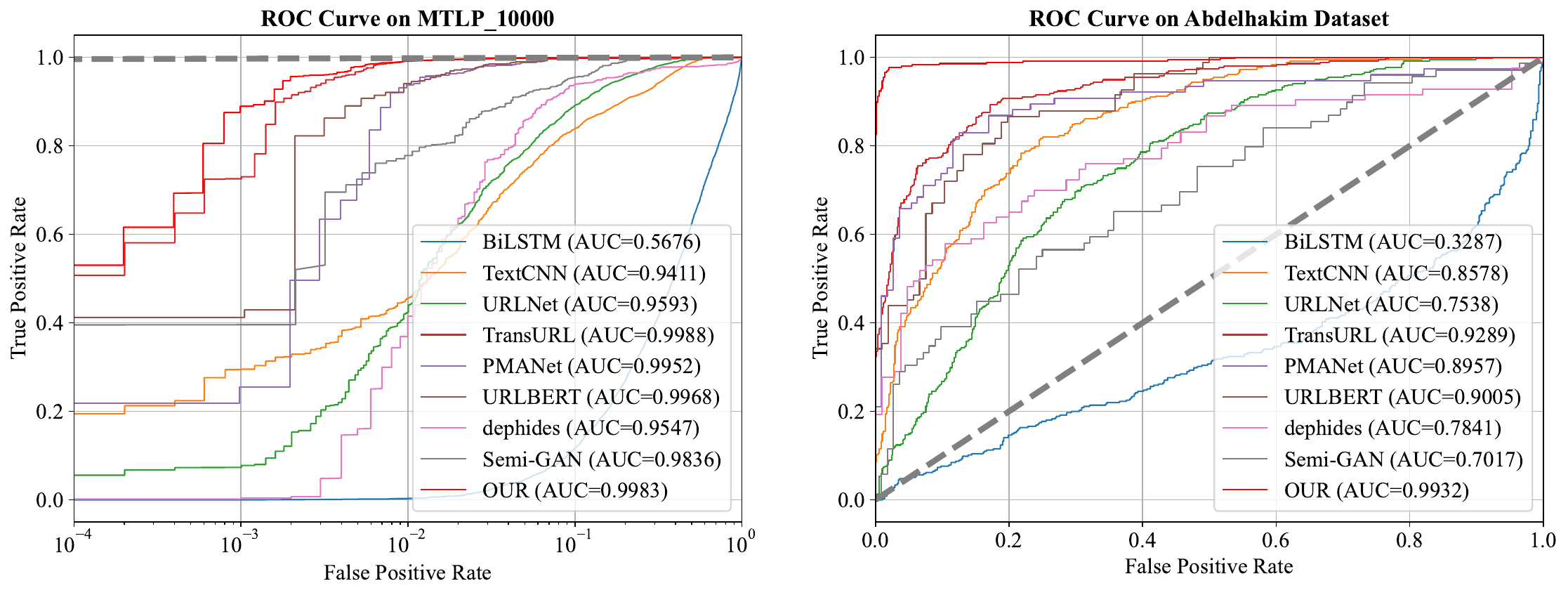}  
    \caption{Performance of our model against other models on the ROC curve.}  
    \label{fig:roc_10000_phishload}  
\end{figure*}

\textbf{Dataset.}
Our experiments focus on using the MTLP Dataset~\cite{10773490} phishing site detection dataset, the Abdelhakim Dataset phishing site detection dataset, and the course-cotrain-data course categorization dataset.

The MTLP dataset is compiled from two different sources: benign samples from the top 2,000 URLs in the Alexa rankings, and additional randomly selected benign and malicious URLs from OpenPhish. The dataset consists of 50,000 benign URLs and 50,000 malicious URLs, which contain HTML content, whois information, and screenshots. Due to the nature of the experiments and equipment limitations, we cleaned the MTLP Dataset and evenly sampled 10,000 data as the training evaluation dataset for this study.

The Abdelhakim Dataset includes 11430 URLs. The dataset are designed to be used as a benchmark for machine learning based phishing detection systems. The datatset is balanced, it containes exactly 50\% phishing and 50\% legitimate URLs. Datasets are constructed on May 2020. The dataset contains a list a URLs together with their DOM tree objects that can be used for replication and experimenting new URL and content-based features overtaking short-time living of phishing web pages.

The course-cotrain-data dataset consists of 1051 pages, with 230 in the course category and 821 in the non-course category.
This data set contains a subset of the WWW-pages collected from computer science departments of various universities in January 1997 by the World Wide Knowledge Base (Web->Kb).


\textbf{Evaluation Metrics.}
To evaluate the model in a more comprehensive and detailed way, we used the following 'TN', 'FP', 'FN', 'TP', 'ACC', 'Precision', 'Recall', 'F1', 'ROC-AUC', 'PR-AUC', 'MCC', 'Weighted F1', 'TPR@FPR=0.0001', 'TPR@FPR=0.001', 'TPR@FPR=0.01', 'TPR@FPR=0.1' assessment metrics.

TP: The number of samples where the model predicts a positive class and the true value is also positive.
TN: Number of samples where the model predicts a negative category and the true value is also negative.
FP: Number of samples where the model predicts a positive class but the true value is negative.
FN: Number of samples where the model predicts a negative class but the true value is positive.
\begin{equation}
ACC = \frac{TP + TN}{TP + TN + FP + FN}
\end{equation}
\begin{equation}
Precision = \frac{TP}{TP + FP}
\end{equation}
\begin{equation}
Recall = Sensitivity = TPR = \frac{TP}{TP + FN}
\end{equation}
\begin{equation}
F1 = \frac{2 \times Precision \times Recall}{Precision + Recall} = \frac{2TP}{2TP + FP + FN} 
\end{equation}

ROC-AUC is the area under the ROC curve, which measures the model's ability to distinguish between positive and negative classes.

PR-AUC is the area under the PR curve, which measures the model's precision at different recall rates.
\begin{equation}
MCC = \frac{TP \times TN - FP \times FN}{\sqrt{(TP + FP)(TP + FN)(TN + FP)(TN + FN)}} 
\end{equation}

If the denominator is zero, the MCC is defined as zero.
\begin{equation}
Weighted\ F1 = \sum_{i=1}^{n} (w_i \times F1_i) 
\end{equation}

where n is the number of categories, $F1_i$ is the F1 score of the $i^{th}$ category, and $w_i$ is the weight of the $i^{th}$ category in the real labeling

TPR@FPR=0.0001, TPR@FPR=0.001, TPR@FPR=0.01, TPR@FPR=0.1: These metrics represent the True Positive Rate (TPR) for a given False Positive Rate (FPR).

\begin{figure*}[t]  
    \centering
    \includegraphics[width=0.95\textwidth]{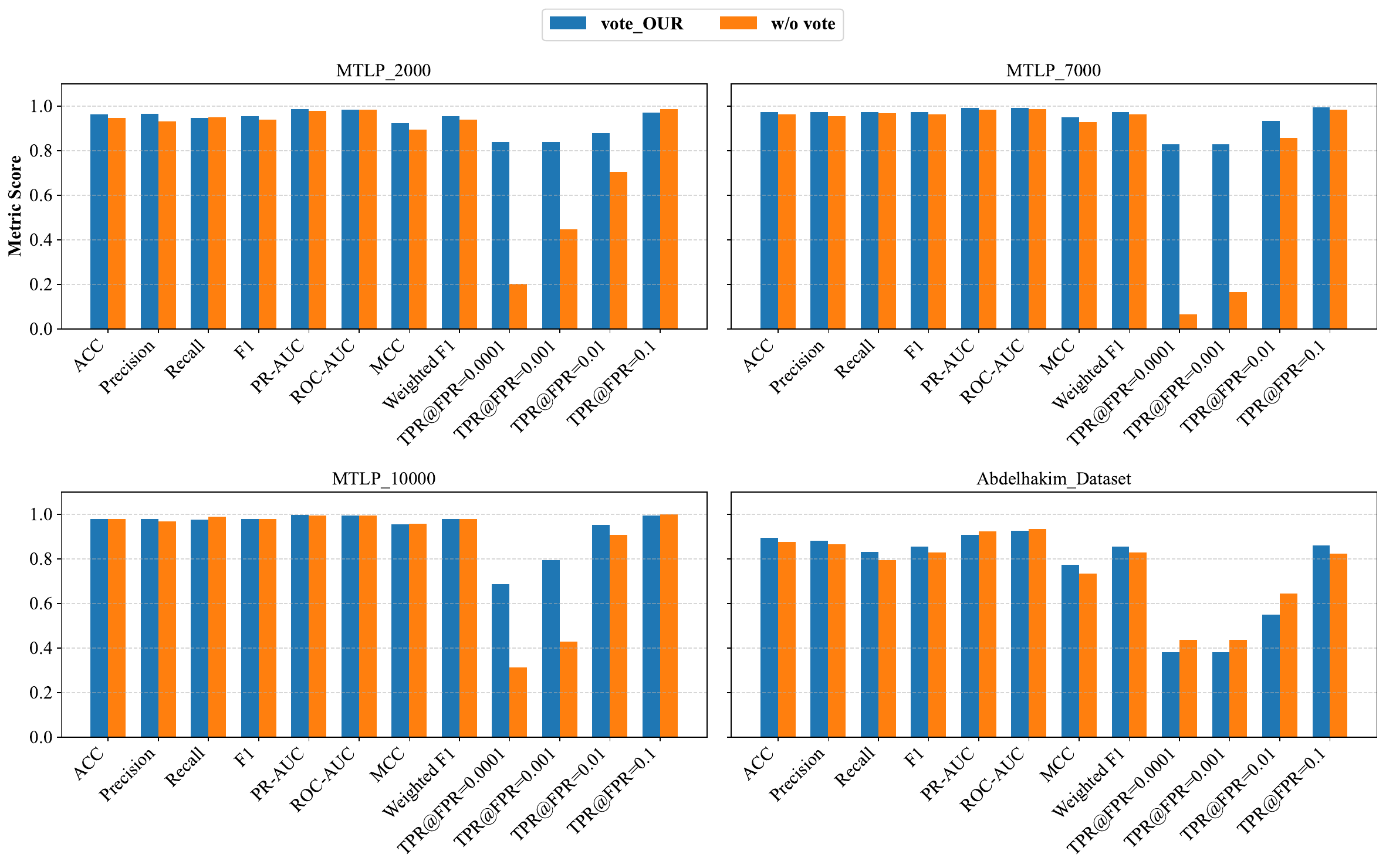}  
    \caption{The ablation experiments of our model against the voting mechanism were done on the MTLP dataset (uniformly sampled 2000, 7000, and 10000 data sizes) and the Abdelhakim dataset, respectively.}  
    \label{fig:vote_unvote}  
\end{figure*}

\textbf{Environment and Parameter setting.}
The batch size during model training was 4, Adam optimizer (initial learning rate: 2e-5, weight decay: 5e-4), dropout rate 0.1, dataset n-fold cross-validation division random seed: 42 and 10 training epochs. We used PyTorch 1.12.1, NVIDIA CUDA12.0 and Python 3.8.20 for training on an NVIDIA 3090.

\subsection{EXPERIMENTAL RESULTS}


\subsubsection{Phishing Detection Capabilities}

In order to comprehensively evaluate the performance of our proposed model WebGuard++ in the field of phishing website detection, we conducted experiments on two publicly available datasets and compared it with several state-of-the-art models. The experimental results include results n the MTLP Dataset with 10,000 data points, (see Table~\ref{tab:10000_1},~\ref{tab:10000_2}), results on Abdelhakim Dataset (see Table~\ref{tab:A_1},~\ref{tab:A_2}), and ROC graphs of different models on both datasets (Figure ~\ref{fig:roc_10000_phishload}).

According to Table~\ref{tab:10000_1},~\ref{tab:10000_2} and the ROC curve on the left side of Figure~\ref{fig:roc_10000_phishload}), our model on the MTLP Dataset with 10,000 data points, has key evaluation metrics such as Accuracy, Precision, F1, MCC, Weighted F1, TPR@FPR(0.0001), TPR@FPR(0.001), TPR@FPR(0.01), and so forth outperformed other models.

Our model achieves the highest values in both ACC and Precision, indicating that it exhibits higher accuracy and both accuracy and false alarm rate in categorizing phishing websites and normal websites. Meanwhile, the optimal values of F1 and MCC further validate the balance and robustness of our model in categorizing positive and negative samples, especially in the case of unbalanced category assignment.
Among them, our model is significantly ahead of other models in low FPR scenarios such as TPR@FPR(0.0001), TPR@FPR(0.001), and TPR@FPR(0.01), which shows a unique advantage in the phishing detection task.

Comparing with other models (see Table~\ref{tab:10000_1},~\ref{tab:10000_2} and Figure~\ref{fig:roc_10000_phishload}), left), traditional models such as BiLSTM and TextCNN have significantly lower performance, indicating their shortcomings in large-scale complex features; whereas advanced models such as PMANet and TransURL outperform our model in F1, MCC, and low FPR metrics, although their AUCs are close. This suggests that our model is not only able to accurately categorize but also provides strong detection capabilities in scenarios such as front and low false alarm rate.

\begin{figure*}[htbp!]  
    \centering
    \includegraphics[width=1\textwidth]{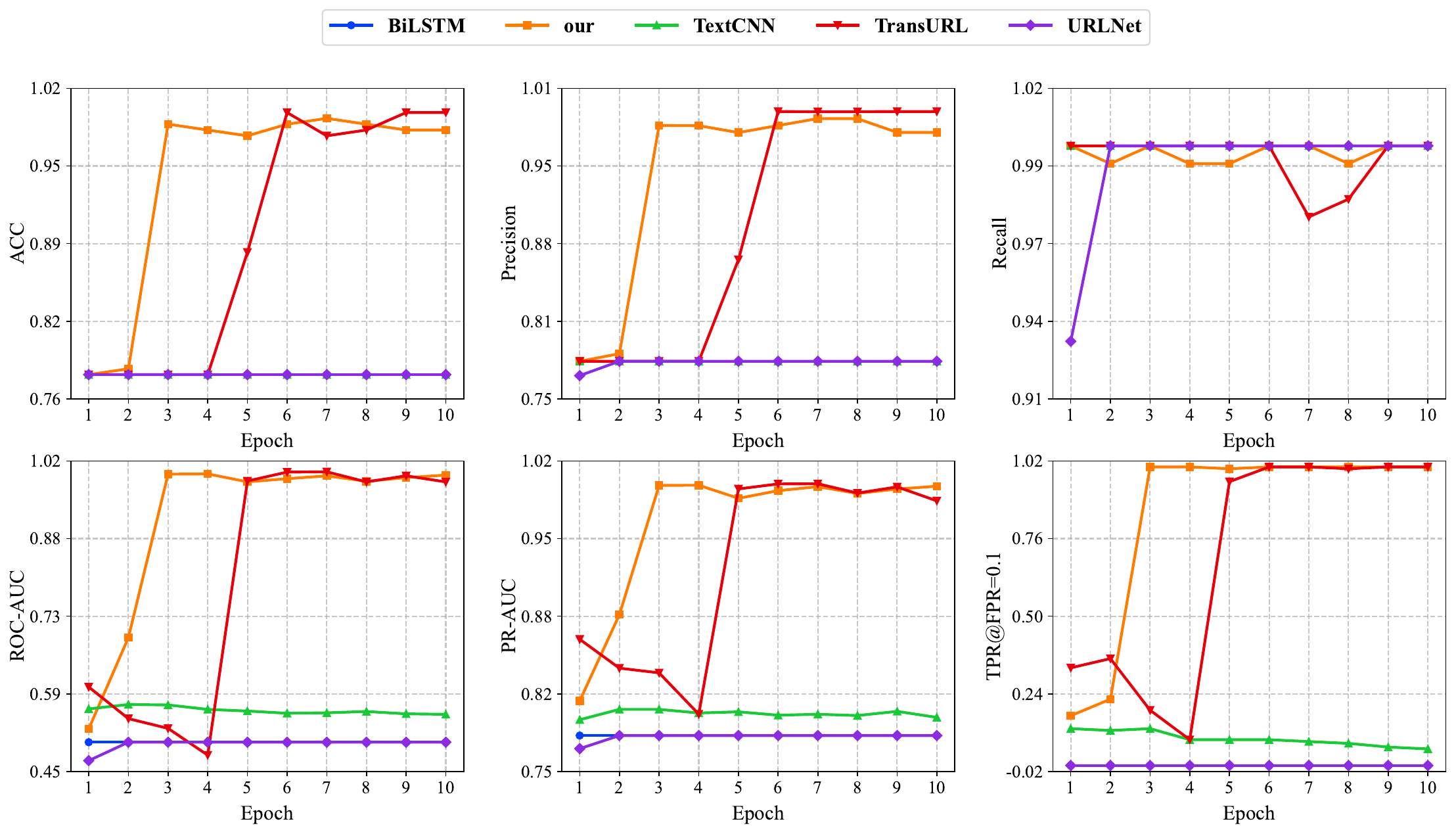}  
    \caption{The generalisation performance of BiLSTM, TextCNN, TransURL, URLNet, and the proposed model in this paper in cross-dataset scenarios is visually demonstrated by the trend changes of six classical evaluation metrics under 10 training epochs.}  
    \label{fig:kua}  
\end{figure*}

According to Table~\ref{tab:A_1},~\ref{tab:A_2} and the ROC curve on the right side of Figure~\ref{fig:roc_10000_phishload}, our model significantly outperforms the other models in ROC-AUC on Abdelhakim Dataset and outperforms the other models in Accuracy, Recall, F1, Weighted F1, PR-AUC, MCC, TPR@FPR(0.0001), TPR@FPR(0.001), TPR@FPR(0.01), TPR@FPR(0.1) and many other key metrics outperform other models. Evidently, our model almost comprehensively overwhelms the other models.

The comparison with other models (see Table~\ref{tab:A_1},~\ref{tab:A_2} and Figure~\ref{fig:roc_10000_phishload} right) shows that the traditional model BiLSTM and the base model TextCNN perform significantly worse on the dataset; whereas models such as PMANet~\cite{LIU2025102638} and URLBERT~\cite{li2024urlbertacontrastiveadversarialpretrained} outperform in some of the metrics, they still do not perform as well as our model in terms of comprehensive performance and low FPR. In addition, some generative models such as Semi-GAN~\cite{kamran2021semi} and dephides~\cite{10388305} perform relatively poorly on Abdelhakim Dataset, which further highlights the strong correlation and generalization ability of our model.

To summarize, our model shows excellent performance in the phishing website detection task, both in terms of classification accuracy, low false alarm rate detection capability, and consistency with unbalanced data, which meets the task requirements.


\subsubsection{Voting Mechanism Ablation Experiment}
In order to verify the actual enhancement effect of the proposed innovative mechanism on model performance, we compare and analyze the performance of adding the voting mechanism with and without adding the voting mechanism under several key evaluation metrics in four data (MTLP\_2000, MTLP\_7000, MTLP\_10000, and Abdelhakim Dataset), as shown in Figure~\ref{fig:vote_unvote}).
From the overall trend, vote\_OUR outperforms w/o vote on most of the assessment metrics, and performs better on most of the key metrics. This result indicates that the innovative mechanism has a significant positive effect in improving the overall recognition performance of the model and enhancing the robustness and generalization ability.
In addition, We observed that after introducing this mechanism, the model's TPR significantly improved at all set FPR thresholds (0.0001, 0.001, 0.01, and 0.1), demonstrating comprehensive and stable detection advantages. This phenomenon indicates that the mechanism can effectively enhance the model's ability to identify positive samples even under extremely low false positive rate requirements, thereby improving the model's practicality and robustness in real-world high-risk scenarios.
In summary, the experiments demonstrate that the proposed innovative method significantly improves the performance of the model under the key indicators while maintaining the overall performance balance, which validates its practical application value in high-reliability task scenarios.As shown in Figure~\ref{fig:vote_unvote}.

\begin{figure*}[t]  
    \centering
    \includegraphics[width=0.95\linewidth]{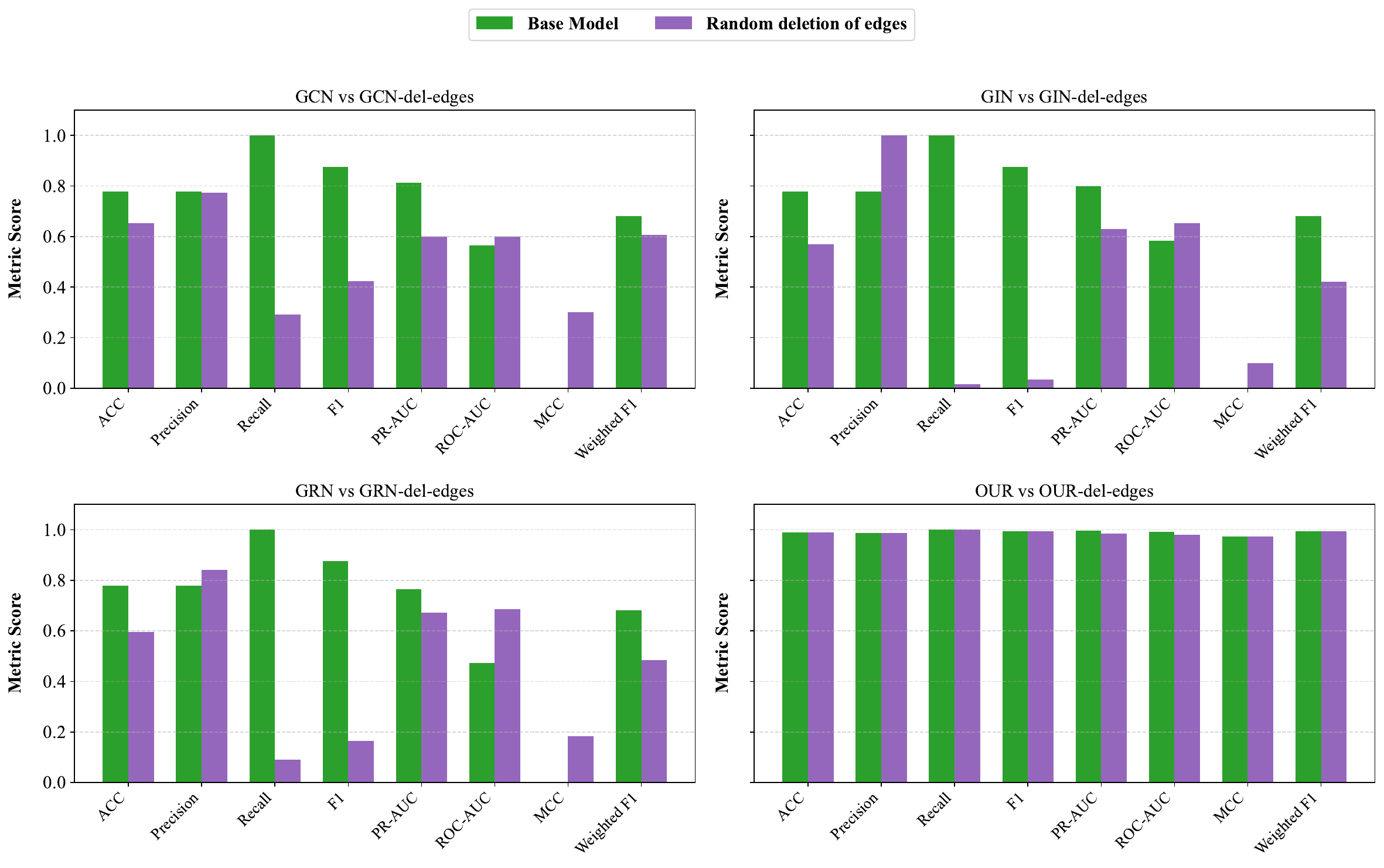}  
    \caption{Compare the robustness test of our model and the basic GNN model in the random edge deletion scenario.}  
    \label{fig:gnn}  
\end{figure*}

\subsubsection{Cross-dataset generalisability test}

We conducted a generalization test on the cross-dataset course\_data, and the results are shown in the Figure~\ref{fig:kua}.
The experimental results presented in the six subplots (a-f) demonstrate the performance of multiple models (BiLSTM, our model, TextCNN, TransTRL, and URLNet) over 10 epochs across six evaluation metrics: Accuracy, Recall, ROC-AUC, PR-AUC, TPR@FPR=0.1 and Precision. Below is a detailed analysis highlighting the advantages of our model compared to the other baselines.

\textbf{(a) Accuracy.}
Our model achieves near-perfect accuracy ($\approx1.0$) after just 2 epochs, significantly outperforming all other models.
TransURL also converges to high accuracy but requires 5 epochs, demonstrating slower convergence.
TextCNN and BiLSTM plateau at lower accuracy values, while URLNet remains constant at a much lower level ($\sim0.8$).
Our model demonstrates faster convergence and higher overall performance, indicating its robustness and efficiency in learning.

\textbf{(b) Recall.}
All models except for URLNet achieve high recall (>0.99) by the end of training. However, our model maintains a consistently high recall throughout the epochs, with minimal fluctuations.
TransURL shows instability during the training process, with notable drops around epochs 5-7.
URLNet lags significantly behind with poor recall ($\sim0.93$).
Our model ensures stability and reliability in recall, a critical metric for minimizing false negatives.

\textbf{(c) ROC-AUC.}
Our model reaches a perfect ROC-AUC of 1.0 within just 2 epochs, outperforming all baselines.
TransURL converges to a similar level but requires 5 epochs, while the other models (TextCNN, BiLSTM, URLNet) fail to surpass 0.75.
URLNet performs the worst, stagnating near 0.5, indicating poor discriminatory capability.
The superior ROC-AUC of our model highlights its exceptional ability to distinguish between classes.

\textbf{(d) PR-AUC.}
Our model achieves a PR-AUC of 1.0 by epoch 2, outperforming all other models in both convergence speed and final performance.
TransURL achieves similar results but requires more epochs (5), while TextCNN and BiLSTM stabilize at much lower levels ($\sim0.8$).
URLNet shows the weakest performance, stabilizing at $\sim0.75$.
The rapid and consistent optimization of PR-AUC demonstrates our model's strength in handling imbalanced datasets by balancing precision and recall.

\textbf{(e) TPR@FPR=0.1.}
Our model achieves a TPR@FPR=0.1 of 1.0 by epoch 2, significantly outperforming all baselines.
TransURL shows delayed convergence, reaching similar performance only after 5 epochs.
TextCNN and BiLSTM remain stagnant at lower values ($\sim0.3$), while URLNet fails to perform effectively, staying near 0.0.
The ability of our model to achieve a perfect TPR at low FPR demonstrates its precision in detecting true positives under stringent conditions.

\textbf{(f) Precision.}
Our model achieves precision values approaching 1.0 by epoch 2, maintaining stability throughout subsequent epochs.
TransURL converges to similar precision levels but requires additional epochs (5).
TextCNN and BiLSTM plateau at much lower precision levels ($\sim0.8$), while URLNet exhibits the poorest performance ($\sim0.75$).
The high precision of our model emphasizes its ability to minimize false positives, an essential property in high-stakes applications.

The experimental results clearly establish the superiority of our model over others. Its faster convergence, higher overall accuracy, and consistent performance across all metrics make it a reliable and efficient solution for the task at hand. Compared to competing models, our approach demonstrates significant advancements in both learning efficiency and classification accuracy, solidifying its position as the state-of-the-art in the given application.

\subsubsection{Robustness Testing}

In the robustness evaluation, we applied random edge deletion with a probability of 50\% to the input graph structures, aiming to assess the resilience of different models under structural perturbations. As illustrated in the bar chart in Figure~\ref{fig:gnn}, our proposed model demonstrates remarkable stability across various evaluation metrics (including ACC, Precision, Recall, F1, PR-AUC, ROC-AUC, MCC, and Weighted F1) with negligible performance degradation and consistently high accuracy.

In contrast, baseline models such as GIN, GRN, and GCN exhibit substantial sensitivity to structural perturbations, with significantly larger performance drops, particularly in Recall and F1 scores. These results, as visualized in Figure~\ref{fig:gnn}, underscore the superior structural robustness and generalization ability of our model, which remains effective even when the graph structure is heavily disrupted.

\section{Conclusion}
In this paper, we propose a novel malicious URL detection framework, WebGuard++, consisting of a cross-scale URL semantic encoder, a subgraph-aware HTML encoder, and a bidirectional multi-view coupling module. The subgraph-aware model allows malicious region signals to aggregate with each other and is less likely to be diluted. At the same time, it also provides ideas for the interpretability of the model, which can be traced to a subregion when maliciousness is detected. In addition, the feature extraction and fusion for multimodal features enable the model to better understand the URL and HTML information and improve the detection performance. In this paper, after extensive experiments, we confirm that our model outperforms the benchmark methods as well as previously proposed state-of-the-art techniques on different data sizes and datasets, and maintains excellent robustness, generalization. Looking ahead, we will endeavor to improve the performance of "WebGuard++" and explore more new approaches.

\section*{Acknowledgment}
This work was supported by the Basic Research Program (No.JCKY2023110C079).

\ifCLASSOPTIONcaptionsoff
  \newpage
\fi

\bibliography{references}

\end{document}